\input harvmac.tex      %% input your version
\input epsf.tex         %%        "
\def\kv{\vec{k}}
\noblackbox
%\input harvmac
%%%%%%%%%%%%%%%%%%%%%%%%%%%%%%%%%%%%%%%%%%%%%%%%%%%%%%%%%%%%%%%%%%%%%
%%%%
%%%%
%%%%        
%%%%             QED for a Fibrillar Medium of Two-Level Atoms 
%%%%                       -Andre' Leclair
%%%%
%%%%
%%%%
%%%%%%%%%%%%%%%%%%%%%%%%%%%%%%%%%%%%%%%%%%%%%%%%%%%%%%%%%%%%%%%%%%%%%
%%%%%%%
%%%%            Macros needed:  harvmac.tex, epsf.tex    (from bulletin board)
%%%%
%%%%
%%%%%%%%%%%%%%%%%%%%%%%%%%%%%%%%%%%%%%%%%%%%%%%%%%%%%%%%%%%%%%%%%%%%%
%%%%%%%%%

%%%%%%%%%%%%%%%%%%%%%%%%%%%%%%%%%%%%%%%%%%%%%%%%%%%%%%%%%%%%%%%
%
%		DEFINITIONS FOR TEX
%
%%%%%%%%%%%%%%%%%%%%%%%%%%%%%%%%%%%%%%%%%%%%%%%%%%%%%%%%%%%%%%%
%
\def\bra#1{{\langle #1 |  }}

\def\tilde{\widetilde}

\def\hat{\widehat}
\def\*{\star}
\def\[{\left[}
\def\]{\right]}
\def\({\left(}		
\def\){\right)}

%
%%%%%%%%%%%%%%%%%%%%%%%%%%%%%%%%%%%%%%%%%%%%%%%%%%%%%%%%%%%%%%%
%

\def\frac#1#2{{#1 \over #2}}
\def\inv#1{{1 \over #1}}

\def\d{\partial}
\def\der#1{{\partial \over \partial #1}}

\def\vev#1{\langle #1 \rangle}
\def\ket#1{ | #1 \rangle}
\def\rvac{\hbox{$\vert 0\rangle$}}
\def\lvac{\hbox{$\langle 0 \vert $}}
\def\2pi{\hbox{$2\pi i$}}

\def\dsl{\raise.15ex\hbox{/}\kern-.57em\partial}
\def\Dsl{\,\raise.15ex\hbox{/}\mkern-.13.5mu D}
%
%%%%%%%%%%%%%%%%%%%%GREEK LETTERS%%%%%%%%%%%%%%%%%%%%%%%%%%%%%%
%

\def\al{\alpha}
\def\ep{\epsilon}

\def\om{\omega}		
\def\sig{\sigma}	

%
%%%%%%%%%%%%%%%%%%%CALIGRAPHIC LETTERS%%%%%%%%%%%%%%%%%%%%%%%%%
%
\def\CA{{\cal A}}

		\def\CL{{\cal L}}
		\def\CO{{\cal O}}

\def\rvac{\hbox{$\vert 0\rangle$}}
\def\lvac{\hbox{$\langle 0 \vert $}}

\def\2pi{\hbox{$2\pi i$}}

\def\dsl{\raise.15ex\hbox{/}\kern-.57em\partial}
\def\Dsl{\,\raise.15ex\hbox{/}\mkern-.13.5mu D}
%
%%%%%%%%%%%%%%%%%%%%GREEK LETTERS%%%%%%%%%%%%%%%%%%%%%%%%%%%%%%
%
%%%%%%%%%%%%%%% MATH CHARACTERS %%%%%%%%%%%%%%%%%%%%%%%%%%%%
%
\font\numbers=cmss12
%\font\numbers=cmu10 scaled\magstep1
\font\upright=cmu10 scaled\magstep1
\def\stroke{\vrule height8pt width0.4pt depth-0.1pt}
\def\topfleck{\vrule height8pt width0.5pt depth-5.9pt}
\def\botfleck{\vrule height2pt width0.5pt depth0.1pt}
\def\Zmath{\vcenter{\hbox{\numbers\rlap{\rlap{Z}\kern
0.8pt\topfleck}\kern
2.2pt
                   \rlap Z\kern 6pt\botfleck\kern 1pt}}}
\def\Qmath{\vcenter{\hbox{\upright\rlap{\rlap{Q}\kern
                   3.8pt\stroke}\phantom{Q}}}}
\def\Nmath{\vcenter{\hbox{\upright\rlap{I}\kern 1.7pt N}}}
\def\Cmath{\vcenter{\hbox{\upright\rlap{\rlap{C}\kern
                   3.8pt\stroke}\phantom{C}}}}
\def\Rmath{\vcenter{\hbox{\upright\rlap{I}\kern 1.7pt R}}}
\def\Z{\ifmmode\Zmath\else$\Zmath$\fi}
\def\Q{\ifmmode\Qmath\else$\Qmath$\fi}
\def\N{\ifmmode\Nmath\else$\Nmath$\fi}
\def\C{\ifmmode\Cmath\else$\Cmath$\fi}
\def\R{\ifmmode\Rmath\else$\Rmath$\fi}
%%%%%%%%%%%%%%%%%%%%%%%%%%%%%%%%%%%%%%%%%%%%%%%%%%%%%%%%%%%%%%%%%
 %%%%%%%%%%%%%%%%%% END OF DEFINITIONS %%%%%%%%%%%%%%%%%%%%%%
 %%%%%%%%%%%%%%%%%%%%%%%%%%%%%%%%%%%%%%%%%%%%%%%%%

\Title{CLNS 96/1407, hep-th/9604100}
{\vbox{\centerline{QED for a Fibrillar Medium of Two-Level Atoms }}}

\bigskip
\bigskip

\centerline{Andr\'e LeClair}
\medskip\centerline{Newman Laboratory}
\centerline{Cornell University}
\centerline{Ithaca, NY  14853}
\bigskip\bigskip

\vskip .3in

We consider a fibrillar medium with a continuous distribution of
two-level atoms coupled to quantized electromagnetic fields. 
Perturbation theory is developed based on the current algebra satisfied
by the atomic operators. 
The one-loop corrections to the dispersion relation for the polaritons
and the dielectric constant are computed.  Renormalization group
equations are derived which demonstrate a screening of the two-level
splitting at higher energies.  Our results are compared with 
known results in the slowly varying envelope and rotating wave approximations.
We also discuss the quantum sine-Gordon theory as an approximate theory. 


\Date{5/96}
%
%
%
%
%
%

%
%
%
%
%sample reference
%
% For the first time you ref an article:
%blah, blah, blah, see the paper\ref\ri{Author, .....} . 
%  The "\ri" is a label of the reference, so that when you
%   reference it later you just write  \ri.  
%
%sample equations
%  
%\eqn\one{
%\V 123 A\rangle_1 \vb_2 \vc_3 = 
%\langle h_1 \left[ V_A (0)\right] h_2 \left[ V_B (0) \right] 
%h_3 \left[ V_C (0) \right] \rangle . }
%
%  Above, \one is a label, so that later in the paper you just
% write  see equation \one.  
%
% Equations with multiple lines:
% \eqn\two{\eqalign{  a & = b \cr c & = d \cr}}
%
%
%  If you want equations  1a, 1b, 1c, separately labeled, etc:
%eq a,b,c etc
%\eqna\three
%$$\eqalignno{
%.............&\three {a} \cr
%..........&\three {b} \cr
%}$$
%\eqn\number{\eqalign{  ......... \cr}}       numbers automatically

\newsec{Introduction }

The interaction of radiation with a medium of atoms is an important
problem with many applications, in particular 
to  the propagation of optical pulses 
and to lasers. 
When studying resonance phenomenon it is natural to approximate
the atoms as two-level systems, resulting in optical Bloch equations. 
For a single atom in an external classical electric field, this problem
was solved by Rabi\ref\rrabi{I. I. Rabi, Phys. Rev. 51 (1937) 652.}.
A collection of atoms coupled to radiation is described by the so-called
Maxwell-Bloch equations.  Propagation effects in the latter model, 
were
studied by McCall and Hahn\ref\rhahn{S. L. McCall and E. L. Hahn, 
Phys. Rev. 183 (1969) 457.}. There, a semi-classical approximation was
made wherein the electromagnetic field was purely classical. 
With some additional approximations (see below) it was shown that
the equations of motion reduce to the well-known sine-Gordon equation,
and the optical  soliton solutions were observed experimentally
also in \rhahn.
The problem  of  
many-atom spontaneous emission was studied in a simplified model
by Dicke\ref\rdicke{R. H.
Dicke, Phys. Rev. 93 (1954) 99.}, which does involve quantized electric
fields.   
We refer the reader to \ref\eberly{L. Allen and J. H. Eberly, 
{\it Optical Resonance and Two-Level Atoms}, Dover Publications, New York, 
1987.} for an excellent account of optical resonance phenomena. 

In this paper we study the fully quantum system of radiation in interaction
with a continuous distribution of atoms in a fiber geometry. 
Two different models are studied, one following from taking the interaction
hamiltonian to be $-\vec{d} \cdot \vec{E}$, which we refer to as
the `current-model', the other following from the minimal coupling
prescription $\vec{p} \to \vec{p} - e \vec{A}$ and referred to as 
the `charge-model'. 
We develop perturbation theory using the current algebra satisfied by
the atomic operators.  This allows us to easily determine the dependence
of various physical quantities on the number of atoms $N$.  In particular
by computing the photon self-energy we determine the first
quantum corrections to the polariton dispersion relation 
and dielectric constant.  

In the current model, 
we show how the quantum corrections imply a renormalization group equation
for the two-level energy splitting, this splitting becoming screened at
higher energies.  

In the last  sections of the paper we compare our results with 
known results obtained under various approximations, namely the 
slowly-varying envelope and rotating wave approximations.  
We also argue that the quantum sine-Gordon theory has some validity
as an effective quantum field theory.  

\def\omo{{\omega_0}}

The two-dimensional quantum field theories we study are interesting
in their own right.  
The more interesting of the two models we study (the current-model)
is defined by the 
hamiltonian
\eqn\hamint{
H = H_0^\phi +  \int dx \( \frac{\omo}{2}  S_3 (x)  + \frac{\beta}{2} 
~ \d_t \phi (x)   \( S^+ (x) 
+ S^- (x) \) \), }
where
$H_0^\phi$ is a free hamiltonian for a scalar field $\phi$, 
$S_3 , S^\pm$ satisfy a current algebra (see equation 2.29),
$\omo$ is the two-level splitting of the atoms, and
$\beta$ is a dimensionless coupling that depends on the 
strength of the dipole transition.  The renormalization group
equation for the radiative level shifts is a consequence of 
the beta-function for $\omo$, which we compute to lowest order. 
To the best of our knowledge, this model is not integrable,
though various approximations to it are integrable (see below).

\newsec{Two Models for the Quantum Maxwell-Bloch Theory}

There are two related models describing the coupling of two-level
atoms to quantized electromagnetic fields.  One follows from taking
the interaction hamiltonian to be $-\vec{d} \cdot \vec{E} $ where
$\vec{d}$ is the dipole moment operator and $\vec{E}$ the electric
field.  The other follows from the usual minimal coupling prescription,
$\vec{p} \to \vec{p} - e \vec{A}$ where $\vec{p}$ is the momentum operator
and $\vec{A}$ the vector potential.  We will  refer to these as the
`current-model' and `charge-model' respectively.  In this section
we describe the reduction of both these models to one spacial dimension.
We will first consider the case of a single atom, and then extend
this to a continuous distribution of atoms.  We set $c=1$ in most
places, but 
keep $\hbar \neq 1$ in some formulas to clarify certain points. 
All formulas with $c\neq 1$ have both $c$ and $\hbar$ restored. 

\def\ev{\vec{E}}
\def\dv{\vec{d}}
\def\om{\omega}
\def\omo{\omega_0}
\def\hatom{H_0^{\rm atom}}
\def\xv{{\vec{x}}}

For both cases, we model a single unperturbed atom as a single 
electron which has two eigenstates 
$\ket{2}, \ket{1} $, with energy difference $\hbar \omega_0$. 
Letting $\hatom$ denote the unperturbed atomic hamiltonian, one has
\eqn\eIIi{
\hatom \ket{1} = - \frac{\hbar \omo}{2} \ket{1} , ~~~~~
\hatom \ket{2} = + \frac{\hbar \omo}{2} \ket{2} . }
In the basis $(\ket{2}, \ket{1} )$, 
\eqn\eIIii{
\hatom = \frac{\hbar \omo}{2} ~ \sigma_3, ~~~~~~
\sigma_3 = \left(\matrix{1&0\cr
0&-1\cr}\right). }

\vfill\eject

\noindent
{\bf {\it 2.1 Current-Model}}
\medskip

One can couple the 2-level atoms to the electromagnetic field 
starting from the energy of an electron in an external electric field.
Let the single atom be centered at $\xv_0$, and let 
$\dv = e \>  \xv$ denote the electric dipole moment operator. 
The complete hamiltonian is 
\eqn\eIIiii{
H = H_0^{\rm field} + \hatom - \dv \cdot \ev (\vec{x}) , }
where $H_0^{\rm field}$ is the free Maxwell hamiltonian, and 
$\vec{x}$ is the position of the electron. 
Let us further assume that the electric field does not vary
significantly over the region where the atomic wavefunction is non-zero,
i.e. 
\eqn\eIIiiib{
\bra{a} \dv \cdot \ev (\vec{x}) \ket{b} \approx \ev (\vec{x}_0 ) 
\bra{a} \dv \ket{b} , }
for $\ket{a,b}$ unperturbed atomic eigenstates. 

The matrix elements of $\dv$ have the following general form
\eqn\eIIiv{\eqalign{
\bra{1} \dv \ket{1} &= \bra{2} \dv \ket{2} = 0 \cr
\bra{2} \dv \ket{1} &= d\, e^{i\alpha} \hat{n} , ~~~~~ 
\bra{1} \dv \ket{2} = d\,  e^{-i\alpha} \hat{n} , \cr}}
where $d$ is a real parameter and $\hat{n}$ is a unit vector that specifies
the orientation of the atom in space.  
$\bra{a} \dv \ket{b} = 0 $ for $a=b$ since $\dv$ is a vector operator
with odd parity and the states $\ket{a}$ are assumed to have definite
parity. 

Letting $\hat{E} = \ev \cdot \hat{n}$, one then has 
\eqn\eIIv{
 \dv \cdot \ev (\xv) = d \, \hat{E}  (\xv_0) \( e^{i\alpha} \sigma^+ 
 + e^{-i\alpha} \sigma^- \) , }
where 
$$ \sigma^+ = \left(\matrix{0&1\cr 0&0\cr}\right) , ~~~~~ 
 \sigma^- = \left(\matrix{0&0\cr 1&0\cr}\right) . $$ 
We will later need the algebra of the $\sig$-operators:
\eqn\eIIvi{
\[ \sig_3 , \sig^\pm \] = \pm 2 \sig^\pm , ~~~~~
\[ \sig^+ , \sig^- \] =  \sig_3 .}
Note that the phase $e^{i\alpha}$ can be removed by letting 
$\sig^\pm \to e^{\mp i \alpha} \sig^\pm$ without affecting the algebra;
we henceforth set this phase to one. 

\def\ahat{\hat{A}}
\def\nhat{\hat{n}}

Consider a fibrillar geometry, where the atom can be viewed as an impurity
in an optical fiber of length $L$ and cross-sectional area $\CA$, where
$L \gg \sqrt{\CA}$.  One can perform a reduction to this essentially
one-dimensional theory as follows.  The free Maxwell action which
determines $H_0^{\rm field}$ is 
\eqn\eIIvii{
S_{\rm Maxwell}  = \inv{4 \pi}  \int d^3 \xv dt \( - \inv{4} 
F_{\mu \nu} F^{\mu \nu} \), }
where $F_{\mu \nu} = \d_\mu A_\nu - \d_\nu A_\mu $.  Let $\hat{x}$ 
denote the direction along the fiber, and $\hat{y}, \hat{z}$ the 
directions transverse to it.  One can first consistently set 
$A_0 = 0$.  We also require the energy flux
to be along the fiber, so that the Pointing vector is in the $\hat{x}$
direction.  This requires $A_x = 0$ and $\d_y A_z = \d_z A_y = 0$. 
Thus, we only have to deal with the components of $\vec{A}$ transverse
to the direction of the fiber.  Of these, only $\hat{A} = \vec{A} \cdot
\hat{n} $ couples to the atom.  Assume $\hat{A}$ is independent of $y,z$,
and let $\int dy dz = \CA$. One has 
\eqn\eIIviii{
S_{\rm Maxwell} = \frac{\CA}{4\pi} \int dx dt \inv{2} 
\( \d_t \ahat \d_t \ahat - \d_x \ahat \d_x \ahat \). }

One can alternatively understand the appearance of the cross-sectional
area $\CA$ by considering the mode expansion of a free scalar field
in finite volume $V$, 
\eqn\finitev{
\Phi (\xv ) = \sum_{\kv} \inv{\sqrt{V}} \inv{\sqrt{2 |\kv|}} 
\( a(\kv ) e^{-i\kv \cdot \xv} + a^\dagger (\kv) e^{i\kv \cdot \xv} \).}
In the fibrillar geometry, $V= \CA L$ and  $\sqrt{\CA}$ is very small 
compared to $L$.  As $L\to \infty$, the modes in the $\hat{x}$-direction
are nearly continuous.  The transverse modes in the $\hat{y}, \hat{z}$
directions have a maximum wavelength on the order of $\sqrt{\CA}$
and are thus very energetic in comparison to the low energy modes in
the $L$-direction.  Thus, we are assuming these high energy transverse
modes are negligibly excited, which is reasonable if 
$1/\sqrt{\CA} \gg \omo$.  In general $\CA$ should be replaced by 
$\CA_{\rm eff}$, which is the effective
cross-sectional area of the fiber as a waveguide. 

Rescale $\ahat$:
\eqn\eIIix{
\ahat = \sqrt{\frac{4 \pi \hbar }{\CA_{\rm eff}} } \phi .}
Then 
\eqn\eIIx{
\frac{S_{\rm Maxwell}}{\hbar} = \int dx dt \inv{2} 
\( \d_t \phi \d_t \phi - \d_x \phi \d_x \phi \). }
The field $\phi$ is a dimensionless scalar field.  In the quantum theory
it satisfies the commutation relations
\eqn\eIIxi{
\[ \phi (x,t) , \d_t \phi (x', t) \] = i \delta (x-x') . }

Using $\hat{E} = - \d_t \hat{A}$, and dividing by $\hbar$ to give the
hamiltonian units of $1/{\rm time}$, we obtain the complete
hamiltonian of the current-model:
\eqn\eIIxii{
H = H_0^{\phi}  + \hatom + H_{\rm int}^{({\rm current})}, }
where $\hatom$ is defined in \eIIii, 
\eqn\eIIxiii{
H_0^{\phi} = \int dx ~ \inv{2} (\d_t \phi)^2  + \inv{2} (\d_x \phi )^2 , }
\eqn\eIIxiiib{
H_{\rm int}^{({\rm current})} 
= \frac{\beta}{2} ~ \d_t \phi (x_0) \( \sig^+ + \sig^- \),
}
and 
\eqn\eIIxiv{
\beta =  \sqrt{ \frac{16 \pi} {\hbar c \CA_{\rm eff} }} \> d . } 
The terminology `current-model' refers to the fact that the
spin operators couple to $\d_t \phi$ which is the spacial component
of the conserved topological current $\ep^{\mu\nu} \d_\nu \phi$, 
where $\ep^{\mu\nu}$ is the anti-symmetric tensor. 

The parameter $\beta$ is the important dimensionless coupling
constant of the model. 
Since $d \sim e R_{\rm atom}$, where $R_{\rm atom}$ is an atomic
dimension, 
\eqn\fine{
\frac{\beta^2}{8\pi} \approx 2 \( \frac{e^2}{\hbar c} \)
\frac{R^2_{\rm atom}}{\CA_{\rm eff}} \approx \frac{2}{137} 
\frac{R^2_{\rm atom}}{\CA_{\rm eff}} .}
Thus, generally, $\beta^2/8\pi$ is very small.  An idealized upper limit
would correspond to a chain of atoms in a waveguide that is one atom
in thickness, so that 
$R_{\rm atom}^2 \sim \CA_{\rm eff}$.  
This is perhaps nearly realizable with a polymer waveguide. 
In this situation $\beta^2 /4\pi \approx 1/137$.
We will refer to this hypothetical limiting case where the quantum
effects are strongest as the quantum optical chain. 

The parameter $\beta^2 /8\pi$ determines the spontaneous decay rate
$1/\tau$ of a single excited atom.  First order perturbation theory
gives
\eqn\eamp{
\left\vert \langle 1,k|2 \rangle \right\vert^2 
= \frac{\beta^2 \omo}{16\pi} \( 2\pi \delta (\omo - |k| ) \)^2 
}
where $|1,k\rangle  = |1\rangle_{\rm atom} \otimes |k\rangle$ and $|k\rangle
$ is a one-photon
state with wave-vector $k$.   This leads to 
\eqn\life{
\inv{\tau} = \int dk 
\left\vert \langle 1,k|2 \rangle \right\vert^2 
(2\pi \delta(\omo - |k|))^{-1} 
= \frac{\beta^2}{8} \omo. }

\vfill\eject

\noindent
{\it 2.2 Charge-Model}
\medskip

\def\av{\vec{A}}
\def\pv{\vec{p}}

For the charge-model we begin with the standard way to couple particles
to electromagnetic fields and consider a single electron hamiltonian
\eqn\eIIxv{
H = \inv{2m_e} \( \pv - e \av (\xv ) \)^2 + V(\xv) , }
where $\pv$ is the momentum operator and $V$ is the atomic potential.
Again, we make a 2-level
approximation for 
 $\hatom = \pv^2 /2m_e + V$,  
and consider states $\ket{1,2}$ as in \eIIi. 
As for the current-model, we assume the analog of \eIIiiib.  Using
\eqn\eIIxvii{
\pv = -i \frac{m_e}{\hbar} \[ \xv , \hatom \] , }
one has 
\eqn\eIIxviii{
- \frac{e}{m_e}  \av (\xv_0) \cdot \bra{a} \> \pv \>
\ket{b} = \frac{i}{\hbar} (E_b - E_a ) 
\vec{A}  (\xv_0)  \cdot \bra{a} \> \dv \>  \ket{b} , } 
where $E_a$ is the energy of the state $\ket{a}$, and $\dv$ is
again the electric dipole operator $\dv = e\>  \xv$. 

Using the parameterization \eIIiv\ for the $\dv$ matrix elements,
and rescaling $\ahat$ as in \eIIix, one obtains a hamiltonian of the
form \eIIxii, where now
\eqn\eho{
H_0^{\phi} = \( \int dx ~ \inv{2} (\d_t \phi)^2  + \inv{2} (\d_x \phi )^2 
\) + \frac{2\pi e^2}{m_e \CA_{\rm eff} } ~ \phi^2 (x_0 ), }
and 
the interaction term is:
\eqn\eIIxix{
H_{\rm int}^{({\rm charge})} = 
-i \frac{\omo \beta}{2}  \phi (x_0) \( \sig^+ - \sig^- 
\) .  }
We refer to this as the `charge-model' since $\phi(\infty) 
-\phi (-\infty) $ is the charge
associated to the topological current $\ep^{\mu\nu} \d_\nu \phi$.

The spontaneous decay rate to lowest order for the charge-model
is the same as for the current-model \life.

\bigskip
\noindent
{\bf {\it 2.3 Continuous Distribution of Atoms}}
\medskip

Consider now a collection of atoms, with the $N$ atoms positioned at
$x=x_i$, $i=1,..,N$.  Also, let $\dv_i = e (\xv - \xv_i )$, 
$\ket{1,2}_i$ and $\sigma (i)$
denote the dipole moment, 2-level states, and Pauli matrix operators
for the i-th atom. 
Since ${}_i \bra{1} \dv_i \ket{1}_i = {}_i \bra{2} \dv_i \ket{1}_i 
= 0$, and ${}_i \langle 1 | 2 \rangle_{i} = 0$, the matrix elements of 
$\dv_i$ don't depend on $\xv_i$.  However, in general the atoms have
variable orientations in space and ${}_i \bra{2} \dv_i \ket{1}_i 
= d \> e^{i\alpha} \nhat_i $, where $\nhat_i$ can vary from atom to 
atom.  To simplify the situation, we assume that all the atoms
are somehow aligned, for example by some external electric field
or by being embedded in a crystal.  Then, 
\eqn\eIIxxi{
{}_i \bra{2} \dv_i \ket{1}_i 
= d \> e^{i\alpha} \nhat, ~~~~~ 
{}_i \bra{1} \dv_i \ket{2}_i 
= d \> e^{-i\alpha} \nhat. }

For the current-model, the interaction becomes 
\eqn\eIIxxii{
H_{\rm int}^{({\rm current})} = \frac{\beta}{2}  \sum_{i=1}^N \d_t \phi (x_i) 
\( \sig^+ (i) + \sig^- (i) \) . }
Introduce space-time dependent spin operators as follows:
\eqn\eIIxxiii{
S^a (x) = \sum_{i=1}^N  \sig^a (i) \delta (x-x_i) . }
These operators satisfy a current algebra
\eqn\eIIxxiv{\eqalign{
\[ S_3 (x,t), S^\pm (x', t) \] &= \pm 2 \> S^\pm (x,t)  \delta(x-x')  \cr
\[ S^+ (x,t), S^-  (x', t) \] &=  S_3 (x,t)  \delta(x-x')  
.\cr}}
The hamiltonian for both the current and charge
models takes the form \eIIxii, where
now 
\eqn\eIIxxv{
\hatom = \frac{\omo}{2}  \int dx  ~ S_3 (x) ,}
and 
\eqn\eIIxxvi{\eqalign{
H_{\rm int}^{({\rm current})} &= \frac{\beta}{2} \int dx ~ 
\d_t \phi (x,t) \( S^+ (x,t) + S^- (x,t) \) \cr
H_{\rm int}^{({\rm charge})} &= -i \frac{\omo \beta}{2} \int dx ~ 
 \phi (x,t) \( S^+ (x,t) - S^- (x,t) \) . \cr}}
For the charge-model, the additional term in \eho\ leads to 
a mass term for the scalar field:
\eqn\ehocont{
H_0^{\phi} =  \int dx ~ \( \inv{2} (\d_t \phi)^2  + \inv{2} (\d_x \phi )^2 
 + \frac{\Delta^2}{2}  ~ \phi^2 (x) \), }
with 
\eqn\eDelta{
\Delta^2 = 4\pi \frac{e^2}{m_e} \frac{N}{L \CA_{\rm eff}} . }

In working with the above formulation, one must impose a further
condition that there is a {\it single} electron bound to each atom.
For the one-atom operators, this is manifest in the two-dimensional
representation of the $\sigma$'s, which has the additional
relations $(\sig^\pm (i) )^2 = 0$, $(\sig_3 (i) )^2 = 1$.  
Note from the definition \eIIxxiii\ that 
\eqn\eIIxxvii{
\( S^\pm (x,t) \)^2 \neq 0. } 
These issues are more easily resolved in a fermionic description,
which we turn to next. 

\vfill\eject

\noindent
{\bf \it 2.4  Fermionic Description}
\medskip

Consider first the one-atom case.  In a second quantized description, one
introduces a fermion wavefunction $\ket{\psi}$:
\eqn\eIIxxviii{
\ket{\psi} = b_1 \ket{1} + b_2 \ket{2} , ~~~~~
\bra{\psi} = \bra{1} b_1^\dagger + \bra{2} b_2^\dagger  , }
where the $b$'s are fermion operators satisfying 
\eqn\eIIxxix{
\{ b_1 , b_1^\dagger \} = \{ b_2 , b_2^\dagger \} = 1. }
The operator $b_1^\dagger$ ($b_2^\dagger$) creates an electron in the
lowest (highest) level.  The $\sigma$ operators then have the following
well-known representation:
\eqn\eIIxxx{
\sig^+ = b_2^\dagger b_1 , ~~~~~
\sig^- = b_1^\dagger b_2 , ~~~~~
\sig_3 = b_2^\dagger b_2 - b_1^\dagger b_1 , }
and the algebra \eIIvi\ is a consequence of \eIIxxix.  Since each atom
has a single electron, one must impose the constraint
\eqn\eIIxxxi{
 b_2^\dagger b_2 + b_1^\dagger b_1 =1. }
 This additional algebraic relation leads to the 2-level relations
$(\sig^\pm)^2 = 0, (\sig_3)^2 = 1$.  

In the multi-atom case, one can define 
\eqn\eIIxxxii{\eqalign{
S^+ (x) &=  b_2^\dagger (x) b_1 (x) , ~~~~~~
S^- (x) =  b_1^\dagger (x) b_2 (x) \cr
S_3 (x) &= b_2^\dagger (x)  b_2 (x)  - b_1^\dagger (x) b_1 (x) . \cr}}
The algebra \eIIxxiv\ is then a consequence of the anti-commutation
relations:
\eqn\eIIxxxiii{
\{ b_1 (x)  , b_1^\dagger (x')  \} = \{ b_2 (x)  , b_2^\dagger (x')  \} 
= \delta (x-x') . } 

Define the number operator:
\eqn\eIIxxxiv{
\hat{N} = \int dx \( 
 b_2^\dagger (x)  b_2 (x)  + b_1^\dagger (x) b_1 (x) \) . }
The operator $\hat{N}$ commutes with the hamiltonian and corresponds to
the number of atoms in the sample.  We therefore impose the constraint
\eqn\eIIxxxv{
\hat{N} = N .}
In evaluating quantum transition amplitudes, one doesn't have to
take special account of this
constraint as long as one deals with initial and final states that
satisfy the constraint, since $[\hat{N}, H] = 0$. 
Note that 
\eqn\eIIxxxvi{
\int dx \> S_3 (x) = - \hat{N} + 2 \int dx \>  b_2^\dagger (x) b_2 (x) . } 

\def\ovac{\ket{ \Omega}}
\def\down{\ket{\downarrow}}

In the sequel, we will work with the perturbative vacuum:
\eqn\eIIxxxvib{
\ovac = \ket{0}_{\rm photon}  \otimes \down_{\rm atom},}
where $\down$ denotes the atomic state with all N atoms in their lowest energy 
state, and $\ket{0}$ is the state with no photons.  This vacuum state
satisfies
\eqn\eIIxxxvic{
b_2 (x) \ovac = S^- (x) \ovac = 0, }
and 
\eqn\eham{
\hatom \ovac = - \frac{N\omo}{2} \> \ovac . }

From \eIIxxxvi, one has
\eqn\eIIxxxvii{
\int dx ~ S_3 (x) ~ \ovac = -N \ovac .} 
Assuming spacial translation invariance of $\ovac$ (we are ignoring boundary
effects for $L$ very large), \eIIxxxvii\ implies
\eqn\eIIxxxviii{
S_3 (x) \ovac = - \frac{N}{L}  \ovac . } 

We remark that $\ovac$ is not the exact ground state of the theory;
see section 5.  

\bigskip

\newsec{Perturbation Theory}

\medskip

\def\lovac{\bra{\Omega}}

In this section, we study the perturbative expansion for the
general correlation functions of the field $\phi$, which
are simply related to electric field correlators via \eIIix.
For convenience of notation, in the charge-model we make the redefinition
$S^\pm \to i^{\pm 1} S^\pm$, which does not affect the commutation
relations \eIIxxiv.  The interaction hamiltonian for both 
models can then be written as 
\eqn\eIIIi{
H_{\rm int} =  g \int dx ~ \CO (x,t) \( S^+ (x,t)  + S^- (x,t) \) }
where for
\eqn\eIIIii{\eqalign{
{\rm current-model}: &~~~~~~ g=\frac{\beta}{2} , ~~~~\CO = \d_t \phi \cr
{\rm charge-model}: &~~~~~~ g=\frac{\omo \beta}{2} , ~~~~\CO =  \phi .
\cr}}

\def\ldown{\bra{\downarrow}}
\def\vac{\ket{0}}
\def\lvac{\bra{0}}

We begin with the partition function, defined as the vacuum to vacuum
transition amplitude.  In the interaction picture, 
\eqn\eIIIiii{
Z = \lovac U(\infty, -\infty) \ovac, }
where 
\eqn\eIIIiv{
U(\infty, -\infty) = T \exp \( -i g \int_{-\infty}^\infty dx dt ~ 
\CO (x,t)  ( S^+ + S^- ) \) .}
The correlator $\ldown S^{a_1} \cdots S^{a_n} \ket{\downarrow}_{g=0}$
is only non-zero when $\sum_i a_i = 0$, thus, 
\eqn\eIIIivb{\eqalign{
Z &= \sum_{n=0}^\infty (-)^n g^{2n} \int dx_{2n} \cdots dx_1 
\int_{t_{2n} > \cdots > t_1 }  dt_{2n} \cdots dt_1 
~ \lvac \CO(t_{2n}, x_{2n} ) \cdots \CO (x_1, t_1 ) \vac_{g=0}
\cr 
& ~~~~~~~~~~\times \sum_{a_1,..,a_{2n} = \pm }
\ldown S^{a_{2n}} (x_{2n} , t_{2n}) \cdots S^{a_1} (x_1 , t_1 ) \down_{g=0}
.\cr}}

\def\kv{\vec{k}}

The time dependence of the S-correlators is simple, since the hamiltonian
at $g=0$ implies 
\eqn\eIIIv{
S^\pm (x,t) = e^{\pm i \omo t } S^\pm (x, 0) . }
Thus, 
\eqn\eIIIvb{
\ldown S^{a_{2n}} (x_{2n} , t_{2n}) \cdots S^{a_1} (x_1 , t_1 ) \down_{g=0}
= \[ \prod_{i=1}^{2n} e^{i a_i \omo t} \]  
\ldown S^{a_{2n}} (x_{2n} , 0) \cdots S^{a_1} (x_1 ,0  ) \down_{g=0}
.}

\def\dkp#1{\frac{d^2 \kv'_{#1}}{(2\pi)^2} }
\def\dk#1{\frac{d^2 \kv_{#1}}{(2\pi)^2} }
\def\gt{{\tilde{G}}}
\def\tn{{2n}}

It will be helpful to pass to momentum space.  Our conventions
are $\vec{k} = (\om, k)$, $d^2 \kv = d\om dk$, $\kv^2 = \om^2 - k^2 $, 
and $ \kv \cdot \xv 
= \om t - k x $. 
Define 
\eqn\eIIIvi{
\lvac \CO (t_{2n}, x_{2n}) \cdots \CO (x_1, t_1 ) \vac_{g=0} 
= \int \dk{1} \cdots \dk{2n} 
\( \prod_{i=1}^{2n} e^{-i \kv_i \cdot \xv_i } \) 
~ \gt_0^{(2n)} ( \kv_1 , \kv_2 ,....., \kv_{2n} ) .}
Substituting this into \eIIIivb, one can perform the time integrals
using 
\eqn\eIIIviii{
\int_{-\infty}^t dt' e^{-i\omega t' }  = \frac{i}{\omega + i\ep} e^{-i\omega
t} , }
where here and below $\ep$ is infinitesimally small and positive.  
One finds 
\eqn\eIIIxi{\eqalign{
Z &= 
\sum_{n=0}^\infty -i g^\tn \int dx_1 \cdots dx_\tn 
\int \dk{1} \cdots \dk{\tn} 
\( \prod_{i=1}^\tn e^{i k_i x_i } \) 
  2\pi \delta \(\sum_i \om_i \)   
 \gt_0^{(\tn)} 
(\kv_1 , ..., \kv_\tn ) 
\cr &~~~~~~~~\times f_{a_1,...,a_\tn}(\om_1 , ..., \om_{\tn -1}
) 
 \sum_{a_1,..,a_{2n} = \pm }
\ldown S^{a_{2n}} (x_{2n} , 0) \cdots S^{a_1} (x_1 , 0 ) \down_{g=0}
, \cr}}
with 
\eqn\eIIIx{
f_{a_1,...,a_\tn}(\om_1 , ..., \om_{\tn -1} ) 
= \prod_{i=1}^{2n-1} \inv{ \sum_{j=1}^i (\om_j - a_j \omo ) + i\ep } . }
The infinite volume singularities in the sum \eIIIxi\ must 
be regulated in order to obtain something meaningful.  However, the
correlation functions, which are obtained by dividing by $Z$,
are well defined order by order in perturbation theory. 

\def\tg{\tilde{G}}

Let $\Phi_1 , \Phi_2 ,...$ denote some local fields.  The same
analysis as above leads to 
\eqn\eIIIxii{
\lovac T \(  \Phi_1 (\xv'_1 ) \cdots \Phi_m (\xv'_m ) \) \ovac  
= \int \dkp{1} \cdots \dkp{m} 
\( \prod_{i=1}^{m} e^{-i \kv'_i \cdot \xv'_i } \) 
~ \gt^{(m)} ( \kv'_1 , \kv'_2 ,....., \kv'_{m} ) ,}
where
\eqn\eIIIxiii{\eqalign{
\gt^{(m)} ( \kv'_1 , ..., \kv'_m ) &= \inv{Z} 
\sum_{n=0}^\infty -i g^\tn \int dx_1 \cdots dx_\tn 
\int \dk{1} \cdots \dk{\tn} 
\( \prod_{i=1}^\tn e^{i k_i x_i } \) 
 2\pi \delta \(\sum_i \om_i \) 
 \cr 
 &~~~~~~~~~\times
\gt_0^{(2n+m)} (\kv'_1 , ...,\kv'_m ; \kv_1 , ..., \kv_\tn ) 
f_{a_1,...,a_\tn}(\om_1 , ..., \om_{\tn -1}
)
\cr &~~~~~~~~~~~~
 \times \sum_{a_1,..,a_{2n} = \pm }
\ldown S^{a_{2n}} (x_{2n} , 0) \cdots S^{a_1} (x_1 , 0 ) \down_{g=0}
, \cr}}
and $\gt_0^{(2n+m)}$ is the Fourier transform of 
\eqn\eIIIxiv{
\lvac T \( \Phi_1 (\xv'_1 ) \cdots \Phi_m (\xv'_n) 
\CO(x_1, t_1) \cdots \CO (x_{2n}, t_{2n} )\) \vac_{g=0} . }
(For $n=0$ in \eIIIxiii\ the $-i$ is omitted.) 

The integrands in \eIIIxiii\ are simple to evaluate.  The 
free Green's functions $\tg_0$ are products of free field propagators
for the field $\phi$.  The S-correlation functions can be evaluated
using the algebra \eIIxxiv\ and \eIIxxxviii.  As usual, dividing
by $Z$ serves to remove `vacuum bubbles'.  We will illustrate the main
features by computing the 2-point function to order $\beta^4$ in the 
next section. 

\vfill\eject

\newsec{Photon Self-Energy and  the Dielectric Constant}

\bigskip

\noindent
{\bf \it 4.1 Charge-Model Computations}
\medskip

Due to the overall momentum conservation, one can write
\eqn\eIVi{
\lovac \phi (x,t) \phi (0) \ovac = \int \frac{d\om dk}{4\pi^2}
~\Pi (\om , k),
} 
where 
\eqn\eIVii{
\Pi (\om , k) = \int \dkp ~ ~\gt^{(2)} (\kv , \kv' ) ,}
and $\gt^{(2)}$ is defined in \eIIIxiii. 
On general grounds, one expects $\Pi$ to take the form 
\eqn\eIVxvii{\eqalign{
\Pi (\om , k) &= \frac{i}{\om^2 - k^2 - \Sigma (w,k)} \cr 
&= \frac{i}{\om^2 - k^2} 
\( 1 + 
\frac{\Sigma}{\om^2 - k^2} + 
\frac{\Sigma^2 }{(\om^2 - k^2)^2 } + \ldots \) . \cr}} 
In quantum electrodynamics, $\Sigma$ is called the photon self-energy. 
The dispersion relation is 
\eqn\edisp{
\om^2 - k^2 - \Sigma = 0.}
In our models, $\Sigma$ is only a function of $\om$, so the dielectric
constant, defined as $\varepsilon (\om) = k^2/\om^2 $, is given by
\eqn\ediel{
\varepsilon (\om ) = 1 - \frac{\Sigma(\om)}{\om^2} . } 

The free field $\phi$ correlators appearing in \eIIIxiii\ are products
of propagators, which follow from the two point function.
For the charge model,
\eqn\eIViii{
\lvac \phi (x,t) \phi (0) \vac_{g=0}  = \int \dk{} ~ 
\frac{i}{\om^2 -k^2 - \Delta^2 + i\ep}
\exp({-i\kv \cdot \xv})
. }
Thus, $\tg_0^{(2n+2)} (\kv, \kv'; \kv_1, ..., \kv_\tn ) $ is a product
of factors $\gt^{(2)}_0 $
where 
\eqn\eIViv{
\gt^{(2)}_0 (\kv_1 , \kv_2 ) = (2\pi)^2 \delta^{(2)} (\kv_1 + \kv_2)
\frac{i}{\kv_1^2 - \Delta^2 + i\ep }. }

For the order $\beta^2$ contribution
one needs
\eqn\eIVv{
\gt^{(4)}_0 (\kv, \kv' ; \kv_1 ,\kv_2 ) =
\gt_0^{(2)} (\kv , \kv') \gt_0^{(2)} (\kv_1 , \kv_2 ) +
\gt_0^{(2)} (\kv , \kv_1) \gt_0^{(2)} (\kv' , \kv_2 ) +
\gt_0^{(2)} (\kv , \kv_2) \gt_0^{(2)} (\kv' , \kv_1 ) . }
In the sum over $a_i$, only $a_1 = + , a_2 = -$ give a non-zero
contribution, and from \eIIxxiv, \eIIxxxviii\ one has
\eqn\eIVvi{
\ldown S^- (x_2, 0) S^+ (x_1 , 0) \down = \frac{N}{L}  \delta(x_2 - x_1 ) .}
One sees that the S-correlation function here gives rise to an effective
2-point interaction of the photons, similar to a mass term. 
The first term in \eIVv\ gives rise to a vacuum bubble which is subtracted.
For the other two terms in \eIVv, one finds that all the integrals are
saturated with $\delta$-functions, and one simply finds
\eqn\eIVvii{
\Pi (\om , k) = \frac{i}{\om^2 - k^2 - \Delta^2 
+ i\ep }  + \frac{ i \Sigma_2 (\om ) }{
(\om^2 - k^2  - \Delta^2 + i\ep)^2 },}
where
\eqn\eIVviib{
\Sigma_2 (\om ) = 
\frac{\beta^2}{2} \frac{N}{L} \frac{\omo^3}{\om^2 - \omo^2 } .}

\def\nl{\frac{N}{L}}

The order $\beta^4$ computation can be carried out explicitly.  
The two S-operator correlators that contribute to the sum are
\eqn\eIVviii{
\ldown S^- (x_4, 0) S^+ (x_3 , 0) S^- (x_2 ,0) S^+ (x_1 , 0) \down 
= \( \frac{N}{L} \)^2 \delta(x_{2 1} ) \delta (x_{43}  ) 
}
\eqn\eIVix{\eqalign{
\ldown S^- (x_4, 0) S^- (x_3 , 0) S^+ (x_2 ,0) S^+ (x_1 , 0) \down 
&= \( \nl \)^2 \( 
\delta (x_{2 3} ) \delta (x_{4 1} ) 
+ \delta (x_{1 3} ) \delta (x_{4 2} ) \) \cr 
&  ~~~ -2 \nl \delta(x_{2 3}) \delta (x_{2 1}) \delta (x_{4 1}) , \cr}}
where $\delta(x_{12}) = \delta(x_1 - x_2)$, etc. 
Doing the $x$ integrals one obtains
\eqn\eIVx{\eqalign{
\Pi (\om , k) \vert_{\beta^4}  &= \frac{-i}{Z} 
\( \frac{\omo \beta}{2} \)^4 
\int \dkp{} \dk{1} \cdots \dk{4}  2\pi \delta(\om_1 + ..+ \om_4 ) 
\gt^{(6)}_0 (\kv, \kv', \kv_1 , ..., \kv_4) \cr
&~~\times \biggl\{  \( \nl \)^2 f_{+-+-} (\om_1 , \om_2 , \om_3 )  
(2\pi)^2 \delta(k_{12} ) \delta(k_{34}) 
+ f_{++--}(\om_1 , \om_2 , \om_3 ) \cr &~~\times \[ \( \nl \)^2 
(2\pi)^2 \( \delta(k_{1 3} ) \delta(k_{2 4}) + \delta(k_{2 3})
\delta(k_{1 4}) \) 
-4\pi  \nl  \delta (k_1 + k_2 + k_3 + k_4 ) \]  \biggr\} 
.\cr}}
($k_{12} = k_1 + k_2$ etc.) 
The free Green's function one needs is
\eqn\eIVxi{\eqalign{
\gt^{(6)}_0 (\kv , \kv'; \kv_1 , \kv_2 , \kv_3 , \kv_4 ) 
&= \gt^{(2)}_0 (\kv, \kv') \( \gt^{(2)}_0 (\kv_1 , \kv_2 ) \gt^{(2)}_0 
(\kv_3, \kv_4 )  + 2 ~{\rm perm. } \)  \cr
&~~~~~+ \( \gt^{(2)}_0 (\kv, \kv_1)  \gt^{(2)}_0 (\kv' , \kv_2 ) \gt^{(2)}_0 
(\kv_3, \kv_4 )  + 11 ~{\rm perm. } \) . \cr}}

\def\ept{\tilde{\ep}}

A diagrammatic technique can be developed for organizing the
computation.   
One first draws a diagram with $n+1$ unconnected lines representing
$\gt^{(2n+2)}_0$, where each line is a propagator assigned the value
$i/(\kv^2 - \Delta^2 + i\ep)$, and one repeats this diagram for each possible
assignment $(\kv, \kv', \kv_1,..., \kv_{2n})$ of the ends of the propagators. 
One then links them according to the $\delta$-functions in the
interactions generated by
the S-correlations, and integrates over all remaining momenta 
including the factors $f(\om)$.  The S-correlation functions are
such that they generate new kinds of interactions at each order,
so we refrain from outlining a complete set of rules here. 

\def\sq{{ \sqrt{k^2 + \Delta^2} } }

For example, the term proportional to $\delta (k_{12} ) \delta(k_{34} )$
leads to the diagram  in figure 1, which equals 
\eqn\eIVxii{
i \( \frac{\omo\beta}{2} \)^4 \( \nl \)^4 
\inv{(\om +\omo) (\kv^2 - \Delta^2 )^2 } 
\int \frac{d\om_2}{2\pi i} ~ 
\inv{(\om_2 - \om + i\ep)(\om_2 - \om_0 + i\ep)
(\om_2^2 - k^2 - \Delta^2 + i\ep)} .}
The labels $1,2..$ in figure 1 refer to $\om_1 , \om_2...$ 
and the structure of the diagram implies $\om_1 = -\om, 
\om_2 = -\om_4 , \om_3 = \om$; this leads to the integrand 
\eIVxii. 
The $\om_2$ integral is easily done.  There are poles at $\om - i\ep$,
$\om_0 - i\ep $, and $\pm (\sq  - i\ept^+ )$.  Closing the contour in
the upper half-plane, one only  picks up the pole at $\om_2 = -\sq$. 
The expression \eIVxii\ becomes
\eqn\eIVxiii{
-i \( \frac{\omo \beta}{2} \)^4 \( \nl \)^2 \inv{2 \sq } 
\( \inv{ (w^2 - k^2 - \Delta^2 )^2  (\om + \omo) (\om + \sq ) 
(\omo + \sq )}  \). }
There are a total of $24$ terms of this type (after subtracting
vacuum bubbles) in \eIVx, all proportional to $(\nl)^2$.

%%%%%%%%%%%%%%%%%%%  fig 1  %%%%%%%%%%%%%%%%%%%%%
\midinsert
\epsfxsize = 2in
\bigskip\bigskip\bigskip\bigskip
\vbox{\vskip -.1in\hbox{\centerline{\epsffile{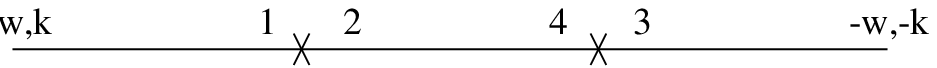}}}
\vskip .1in
{\leftskip .5in \rightskip .5in \noindent \ninerm \baselineskip=10pt
Figure 1. Diagrammatic representation of \eIVxii. 
\smallskip}} \bigskip
\endinsert
%%%%%%%%%%%%%%%%%%% End fig 1   %%%%%%%%%%%%%%%%%%%%%%%

The term in \eIVx\ proportional to $\delta(k_1 + k_2 + k_3 + k_4)$
is a new effective 4-point interaction of the photons. In the
diagrammatic scheme described above, one contribution for example
is shown in figure 2.   This has the value
\eqn\eIVxiv{
\inv{8} \nl (\omo \beta)^4 \inv{2\om_0 (\om + \omo)(\om^2 - k^2 - \Delta^2 
)^2 }
\int \dk{3} \inv{ (\kv_3^2 - \Delta^2 + i\ep )(\om_3 - \omo + i\ep)} 
. }
There are a total of 12 terms of this kind.  

\def\omt{\tilde{\om}}
\def\kt{\tilde{k}}

%%%%%%%%%%%%%%%%%%%  fig 2  %%%%%%%%%%%%%%%%%%%%%
\midinsert
\epsfxsize = 1.5in
\bigskip\bigskip\bigskip\bigskip
\vbox{\vskip -.1in\hbox{\centerline{\epsffile{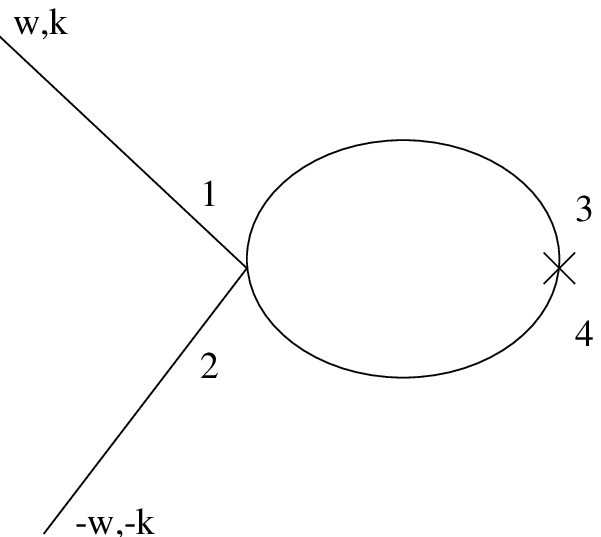}}}
\vskip .1in
{\leftskip .5in \rightskip .5in \noindent \ninerm \baselineskip=10pt
Figure 2.  Diagrammatic representation of \eIVxiv. 
\smallskip}} \bigskip
\endinsert
%%%%%%%%%%%%%%%%%%% End fig 2   %%%%%%%%%%%%%%%%%%%%%%%

Upon summing these contributions, one finds
\eqn\eIVxv{
\Pi (\om, k) \vert_{\beta^4} 
= \frac{i}{(\om^2 - k^2 - \Delta^2 )^3} (\Sigma_2 (\om) )^2 
+ \frac{i}{(\om^2 - k^2 - \Delta^2 )^2}  \Sigma_4 (\om ) }
where 
\eqn\eIVxvi{\eqalign{
\Sigma_4 (\om )&=  -\frac{i}{4} \nl \beta^4 \omo^5 \int 
\frac{d\omt d\kt}{(2\pi)^2} 
\inv{\omt^2 - \kt^2 - \Delta^2 + i\ep} 
\biggl[ \frac{1}{(\om^2 - \omo^2) 
 (\omt - \omo + i\ep)^2 } \cr
 &~~~~~~~
+  \frac{2 \omo}{(\om^2 - \omo^2)^2  
 (\omt - \omo + i\ep) } \biggr]  .\cr }}
The first term in \eIVxv\ arises from the sum of all diagrams of the
kind shown in figure 1, whereas $\Sigma_4$ is the sum of  the diagrams
of the kind shown in figure 2.   Thus we have verified to fourth order
the structure anticipated in \eIVxvii, with $\Sigma = 
\Delta^2 +  \Sigma_2 + \Sigma_4
+ \CO(\beta^6)$. 

In theories with more conventional perturbative expansions, the
self-energy is the sum of one-particle irreducible 2-point diagrams.  
One-particle irreducibility is likely to be more transparent
in a perturbative expansion based on the $b_{1,2}$ fermion fields
rather than directly on the $S$ operators, but we don't develop this
here.

The left-over integrals in $\Sigma_4$ identify it as a one-loop
contribution.  
The final result for the self-energy of the charge-model is the following:
\eqn\eIVxxiii{\eqalign{
\Sigma(\om) =  \Delta^2 + \frac{\beta^2}{2} \nl &
\frac{\omo^3}{\om^2 - \omo^2} 
 \biggl( 1 + \frac{\beta^2}{8\pi}  \biggl[  
\frac{2 \omo^2}{\omo^2 - \Delta^2}  
- \frac{2 \omo^3}{\sqrt{\omo^2 - \Delta^2}} 
\cr
&\times 
\frac{ 3\omo^2 - \om^2 - 2 \Delta^2 }{(\om^2 - \omo^2)(\omo^2 - \Delta^2 )} 
\( \log \( \frac{ \omo - \sqrt{\omo^2 - \Delta^2 } }{\Delta} \) - 
\frac{i\pi}{2}  \)  
\biggr]  \biggl) 
. \cr}}
The size of the one-loop corrections is determined by the dimensionless
parameter $\beta^2/8\pi$.  
The expression \eIVxxiii\ is the main result of this section. 
We remark that setting $\Delta$ to zero leads to infra-red divergences.

The self-energy depends on $N$ only through the combination 
$\beta^2 N \omo/L$ with units of ${\rm mass}^2$.  This has a universal
meaning in terms of the volume density $\eta $ of atoms:
\eqn\eIVxxiiib{
m^2  \equiv \frac{\beta^2}{8} \nl \omo = \inv{\tau} \nl = 
2 \pi \hbar \omo 
 \frac{d^2 }{c^4} \eta  
, }
with 
$\eta =  \frac{N}{\CA_{\rm eff} L} $.  
In the limiting case of the quantum optical chain discussed above,
\eqn\chain{
m^2 \sim \inv{137} ~ 2 \pi \omo ~ \rho,}
where $\rho$ is density per unit length of atomic impurities. 
To obtain an idea of orders of magnitude, for $\hbar \omo = 1 {\rm ev}$,
one finds $m = 1 {\rm ev}$ for $\rho = 10^5 /{\rm cm}$.  

\bigskip
\noindent
{\bf \it 4.2 Current-Model Computations}
\medskip

Consider next the current-model.  The photon self-energy should
now be defined as follows:
\eqn\eIVxxiv{
\lovac \d_t \phi (x,t) \> \d_t \phi (0) \ovac = i
\int \dk{}~ \exp({-i\kv \cdot \xv })  \frac{\om^2}{\om^2 - k^2 - \Sigma} .}
The general 
expressions for the charge-model we obtained in this section still
apply, except that now $\gt^{(n)}_0 (\kv_1 , ..., \kv_n )$ is the 
Fourier transform of the correlation function 
$\lvac \d_t \phi (\xv_1 ) \cdots \d_t \phi (\xv_n ) \vac_{g = 0}$.  
In particular, \eIViv\ is replaced by 
\eqn\eIVxxv{
\gt^{(2)}_0  (\kv_1 , \kv_2 ) = (2\pi)^2 \delta^{(2)} (\kv_1 + \kv_2 ) 
\frac{i \om_1^2 }{\kv_1^2 + i\ep} . } 
Repeating the above computations, one finds
\eqn\eIVxxvi{\eqalign{
\Sigma (\om ) &=  \frac{\beta^2}{2} 
\nl  \frac{ \omo \om^2}{\om^2 - \omo^2}  
-\frac{i}{4} \nl \beta^4 \om^2 \omo  \int \frac{d\omt d\kt}{(2\pi)^2} 
 \frac{ \omt^2}{\omt^2 - \kt^2 + i\ep} 
\biggl[ \frac{1}{(\om^2 - \omo^2) (\omt - \omo + i\ep)^2 } 
\cr &~~~~~~~~~~
+ \frac{2 \omo}{(\om^2 - \omo^2)^2 (\omt - \omo + i\ep)} \biggr] 
.\cr }} 

One can perform the integrals as before.  An important difference
from the charge-model is that here the one-loop 
integrals are ultraviolet 
divergent.  There is a natural u.v. cutoff in the
model since it is implicit in \eIIiiib\ that the 
wavelengths of the photons are large compared to the size of
the atoms.  We introduce a u.v. cutoff $\mu$ as follows:
\eqn\eIVxxvii{
\int_0^\infty d \om  \to \int_0^\mu d\om .} 
The final result is 
\eqn\eIVxxviii{
\Sigma(\om) = \frac{\beta^2}{2} \nl   \frac{\omo \om^2 }{\om^2 - \omo^2} 
\[ 1 +  \frac{\beta^2}{4\pi}   
- 
\frac{\beta^2}{8\pi}    \( \frac{\om^2 + \omo^2 }{\om^2 - \omo^2}\) 
\( \log \( \frac{ \mu^2 }{\omo^2 }  - 1 \) - i\pi \) \]  
 ~~~~~+ \CO( \beta^6) .} 
Here, the imaginary part arises from choosing a positive argument of the
$\log$, i.e. $\mu^2 > \omo^2$, which is physically sensible if 
resonant photons (with energy $\omo$) have a longer wavelength than
the size of atoms.  Note however that ${\rm Im}(\Sigma) = 0$ 
when $\mu^2 < \omo^2 $.  
As we show below, 
the occurrence of this imaginary part is related to the spontaneous
decay lifetime of the atoms. 

\bigskip
\newsec{Polaritons and Spontaneous Decay}

One sees from the self-energy that the spectrum consists of two
`polariton' branches.  To lowest order in $\beta^2$, the dispersion
relation for the current-model reads
\eqn\equasi{
\om^2 = \om^2_\pm  = \inv{2} \( ( \omo^2 + k^2 + 4 m^2)  \pm 
\sqrt{ (\omo^2 - k^2 - 4 m^2 )^2 + 16 m^2 \omo^2 } \).}
These two branches are plotted in figure 3. 
As $m \to 0$, the two branches become $\om^2 = \omo^2$ and
$\om^2 = k^2$, i.e. optical phonon-like and photon-like respectively. 
For finite $m$, the $\om_+$ branch is phonon-like for small $|k|$
but photon-like for large $|k|$, and visa versa for the $\om_-$
branch.  
In general one has quasi-particles with both atomic and photon
degrees of freedom.  
%\eqn\eIVxxvi{
%\om^2 = \om_\pm^2 =  \inv{2} \( ( \omo^2 + k^2 )\pm 
%\sqrt{ (\omo^2 - k^2)^2 + 16 m^2 \omo^2 } \).}
One has 
$\om_+ (k=0) \approx \omo + 2m^2/\omo$, and 
$\om_- (k=\infty) \approx \omo$.  Thus there is a gap between
the two branches with
\eqn\egap{
E_{\rm gap} = \frac{2 m^2}{\omo} = \frac{\beta^2}{4} \nl = 4\pi 
d^2 \eta.}  
For the idealized quantum optical chain with $\beta^2 /4 \pi 
\approx 1/137$, if $N/L = 10^5 /{\rm cm} $, 
then $E_{\rm gap} = 2 {\rm ev}$.  

The analagous formula for the charge-model is 
\eqn\eIVxxvi{
 \om_\pm^2 =  \inv{2} \( ( \omo^2 + k^2 + \Delta^2  )\pm 
\sqrt{ (\omo^2 - k^2 - \Delta^2 )^2 + 16 m^2 \omo^2 } \).}
For $\Delta^2/\omo^2 , ~ m^2/\omo^2 \ll 1$, the gap between
the two branches is the same as in \egap.   The main difference
between the charge and current models is for the $\om_-$ branch
near $k=0$.  One finds $\om_-^2 (k=0) \approx \Delta^2 - 4 m^2$. 
For $\Delta^2 = 4 m^2$, $\om_- = |k|$ for $k \approx 0$, and
the dispersion relations \eIVxxvi\ and \equasi\ are actually
identical.
In \ref\rkon{R. Konik and A. LeClair, in preparation.}\ the
analagous problem with harmonic oscillator defects rather than
two-level atoms is studied, and there it is shown that the analog
of the cancelation $\Delta^2 = 4 m^2$ occurs.  This justifies
enforcing $\Delta^2 = 4 m^2$ in order to obtain a physical
photon dispersion relation near $k=0$.

%%%%%%%%%%%%%%%%%%%  fig 3  %%%%%%%%%%%%%%%%%%%%%
\midinsert
\epsfxsize = 2.5in
\bigskip\bigskip\bigskip\bigskip
\vbox{\vskip -.1in\hbox{\centerline{\epsffile{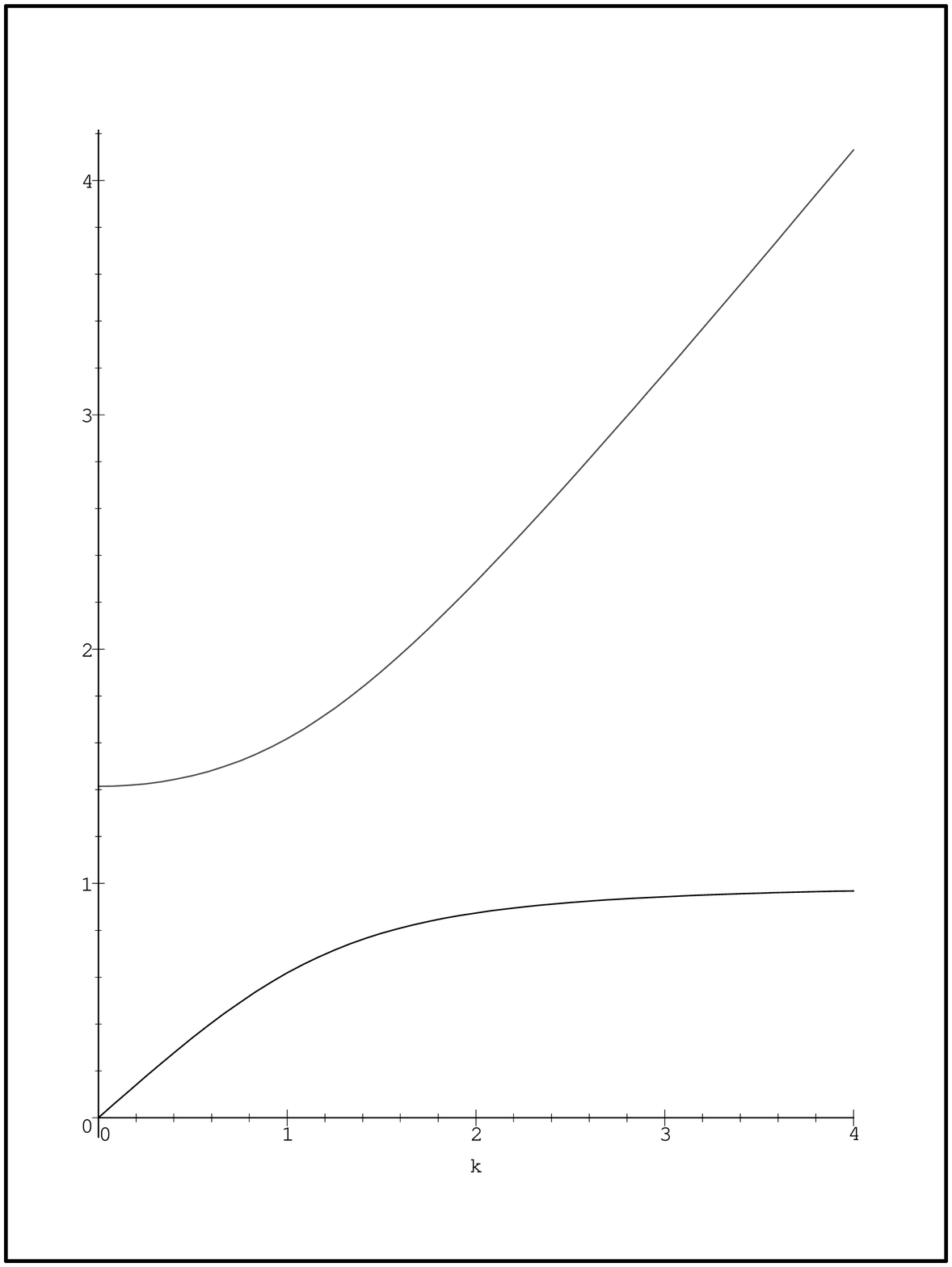}}}
\vskip .1in
{\leftskip .5in \rightskip .5in \noindent \ninerm \baselineskip=10pt
Figure 3.
The dispersion relation $\om$ verses $k$ for the current-model. 
(For  $\omo=1$, $m= .5$.)
\smallskip}} \bigskip
\endinsert
%%%%%%%%%%%%%%%%%%% End fig 3   %%%%%%%%%%%%%%%%%%%%%%%

Due to the imaginary part of $\Sigma$, and consequently the imaginary
part of the dielectric constant, there is attenuation of plane waves
in the medium.  
The origin of this imaginary part is the finite
lifetime under spontaneous decay  of  single  excited atomic states. 
To see this, note that for a small shift of  $\omo$ 
\eqn\edelom{
\frac{ \omo + \delta \omo}{\om^2 - (\omo + \delta\omo)^2} 
\approx 
\frac{\omo}{\om^2 - \omo^2} 
\( 1 + \frac{\delta\omo}{\omo} ~\frac{\om^2 + \omo^2}{\om^2 - \omo^2} \).}
Comparing this with \eIVxxviii, one sees that the imaginary part of
$\Sigma$ for the current-model can be interpreted as a small
imaginary shift to $\omo$:
\eqn\eshift{
\frac{\delta \omo}{\omo} = i \frac{\beta^2}{8} . }
This leads to the decay of single atoms $e^{i\omo t} \to 
e^{i\omo t} e^{-t/\tau} $, with $\tau$ given in \life. 

The attenuation constant $\alpha$ is defined such that
the intensity of radiation decays as $e^{-\al x}$.  In terms of the
dielectric constant, 
\eqn\eIVxxiv{
\al = \left\vert \om \frac{ {\rm Im} 
\varepsilon }{\sqrt{{\rm Re} \varepsilon }} 
\right\vert .}
For the current-model,  one finds,  
\eqn\alE{
\alpha \approx  4 \pi m^2 \( \frac{\beta^2}{8\pi} \) 
\frac{ \om^2 + \omo^2 }{(\om^2 - \omo^2)^2 } 
\vert \om \vert . }

\bigskip
\newsec{Renormalization Group for the 2-Level Splitting}
\medskip

We have seen that in the current-model the photon self-energy depends on
an ultraviolet cutoff $\mu$.  A basic idea of the renormalization
group is that couplings  also depend on the scale $\mu$ 
in such a way that physical quantities are independent of $\mu$. 
The result \eIVxxviii\ has precisely a form that allows us to 
consider $\omo$ as a function of $\mu$, since 
\eqn\ederiv{
\d_{\omo} \( \frac{\omo}{\om^2 - \omo^2} \) = \frac{\om^2 + \omo^2}{(\om^2
- \omo^2)^2} . }
The $\mu$-independence of $\Sigma$ amounts to the renormalization
group equation
\eqn\eVii{
\mu \frac{d}{d\mu} \Sigma = 
\( \mu \der{\mu}  + \( \mu \der{\mu} \omo \) \der{\omo} \) \Sigma = 0, }
or equivalently
\eqn\eViii{
\Sigma (\om , \omo (\mu_2 ) , \mu_2 )
= \Sigma (\om , \omo (\mu_1 ) , \mu_1 ) .} 
To order $\beta^2$, \eVii\ leads to the beta-function:
\eqn\eViv{
\mu \der{\mu} \omo = - \frac{\beta^2}{4\pi}  \omo . }
Since \eVii\ could be satisfied assuming no dependence of $\beta$
on $\mu$ we conclude  that up to order $\beta^4$, $\beta$ is unrenormalized. 

The beta-function \eViv\ means that the parameter $\omo$, which has
engineering dimension $1/{\rm time}$, has an additional anomalous dimension 
of $-\beta^2 /4\pi $.  Integrating \eViv, 
\eqn\eVv{
\frac{ \omo(\mu_2 ) }{\omo (\mu_1 )} 
= \( \frac{\mu_1}{\mu_2} \)^{\beta^2/4\pi} . }
Thus, as $\mu$ increases, $\omo$ decreases, reaching an ultraviolet
fixed point at $\omo =0$.  

One can derive a relation which describes the behavior of $\Sigma$ 
as one scales the dimensionful parameters $\om, \omo, L$.  Ordinary
dimensional analysis implies
\eqn\eVvi{
\Sigma (\om , \omo (\mu), L, \mu ) 
= \mu^2 \tilde{\Sigma} \( \frac{\om}{\mu} , \frac{\omo(\mu)}\mu ,
L\mu \), }   
where $\tilde{\Sigma}$ is a function of dimensionless parameters. 
Rescaling all dimensionful parameters by a dimensionless
parameter $s$, and using the renormalization group equation 
\eViii, one has 
\eqn\eVvii{
\mu^2 \tilde{\Sigma} \(\frac{s\om}\mu  , \frac{s\omo (\mu)}\mu , L\mu /s \) 
= \mu'^2 \tilde{\Sigma} \( \frac{s \om}{\mu'} , \frac{s \omo(\mu')}\mu' ,
L\mu'/s  \). }   
Taking $\mu' = s \mu$, one obtains the scaling equation 
\eqn\eVviii{
\Sigma (s \om , s \omo (\mu ) , L/s , \mu )
= s^2 \Sigma (\om , \omo (s \mu ) , L, \mu ) .} 
This means that at higher energies $s\om$, the two-level splitting
$\omo (\mu)$ is screened to $\omo (s\mu) =  s^{-\beta^2/4\pi} \omo (\mu)$. 

\bigskip
\newsec{Comparison with the Reduced Maxwell-Bloch Theory} 

\medskip

Two approximations commonly made in the quantum optics literature
are the so-called slowly varying envelope and rotating wave approximations.
In this section we compare the above results with the analagous
results obtained in these approximations.  
\medskip
\noindent
{\it  6.1 Definition of the Model}
\medskip

\def\dk{ \frac{dk}{\sqrt{2\pi}} }
\def\dke{ \frac{dk_e }{\sqrt{2\pi}} }
\def\ah{\hat{a}}

The slowly varying envelope approximation is suitable for dealing
with near resonant phenomena.  Consider the mode expansion of the
free scalar photon field:
\eqn\eVIi{
\phi (x,t) = \int \dk \inv{ \sqrt{2 |k|}}  
\( a(k) e^{-i \kv \cdot \xv} + a^\dagger (k) e^{i\kv\cdot \xv} \), }
where $\kv = (|k|, k)$. 
The photon creation operators 
satisfy
\eqn\eVIii{
[ a(k) , a^\dagger (k') ] = \delta (k-k') ,}
which implies \eIIxi.  
We suppose that near resonant photons are most important, 
and let 
\eqn\eVIiii{
k = \omo + k_e , }
where $k_e$ denotes  an `envelope' wave vector.  Letting $|k| \approx \omo$, one
has 
\eqn\eVIiv{
\phi (x,t) \approx
e^{-i\omo(t-x)} \psi (x,t)  + e^{i\omo(t-x)}  \psi^\dagger (x,t),}
where
\eqn\eVIv{\eqalign{
\psi (x, 0)  &= 
\inv{\sqrt{2\omo}}  \int \dke  ~ \hat{a}(k_e) e^{ik_e x}  \cr
\psi^\dagger (x, 0)   &= 
\inv{\sqrt{2\omo}}  \int \dke  ~ \hat{a}^\dagger (k_e) e^{-ik_e x}  , \cr}}
and \eqn\eVIvi{
\hat{a} (k_e) =  a(\omo + k_e). }
The operators $\ah, \ah^\dagger$ satisfy the same commutations as
\eVIii.  

In classical theory, $\psi$ and $\psi^\dagger$ are referred to as the
slowly varying envelopes if $k_e^2 \ll \omo^2 $, which implies 
\eqn\eVIvii{
|\d_x \psi | \ll \omo | \psi | , 
~~~~~|\d_t \psi | \ll \omo | \psi |.}  

Note that in expanding the field as in \eVIiv, we are quantizing
about a {\it right-moving} plane wave.  One can also begin with 
envelopes of left-moving waves separately; however we will not
consider interactions between the left and right moving envelopes. 

Using \eVIvii, one finds 
\eqn\eVIviii{
\int dx dt \inv{2} \( (\d_t \phi)^2 - (\d_x \phi)^2 \) 
\approx 2 i \omo \int dx dt ~ \psi^\dagger (\d_x + \d_t ) \psi .}
Without any additional interactions with atoms, the equation of motion
is $(\d_x + \d_t ) \psi = 0$, which means that the envelope also consists
of right-moving excitations only at zero coupling.  
The canonical commutation relation which follow from 
\eVIviii\ are 
\eqn\eVIix{
\[ \psi (x,t), \psi^\dagger (x', t) \] = \inv{2\omo} \delta (x-x'),}
and is compatible with \eVIv.  

The interaction \eIIxxvi\ contains terms with both photon
creation operators $a^\dagger $  and $S^+$ operators which excite the atoms. 
Such terms lead to vacuum fluctuations wherein both photons 
and atoms are simultaneously excited, and also to real processes
where e.g. an incoming photon excites the atom and emerges as two photons. 
The rotating wave approximation sets such processes to zero.  
For the charge-model, the rotating wave approximation, combined with
the slowly varying envelope approximation leads to 
\eqn\eVIx{
H_{\rm int} = -i \frac{\omo \beta}{2}  \int dx 
\( \psi e^{-i\omo (t-x)} S^+ - \psi^\dagger e^{i\omo (t-x)} S^- \).}
Note that the phases $e^{\pm i \omo t}$ in 
\eVIx\ cancel the time dependence \eIIIv\ of $S^\pm$ which comes from
the hamiltonian $\hatom$.  Thus we can replace 
$e^{\mp i\omo (t-x)} S^\pm$ with new operators $\hat{S}^\pm$ 
and set the $\hatom$ piece of the hamiltonian to zero.  Thus
we consider the model defined by the complete hamiltonian:
\eqn\eVIxi{
H = - 2 i \omo \int dx ~ \psi^\dagger \d_x \psi  
-i \frac{\omo \beta}{2} \int dx \( \psi S^+ - \psi^\dagger S^- \) .}
The first term follows from \eVIviii\ and we relabeled the 
$\hat{S}^\pm$ operators back to $S^\pm$.  The algebra satisfied
by the $S$'s is the same as before.  
In the classical context, the equations of motion for the model 
\eVIxi\ is sometimes referred to as the reduced Maxwell-Bloch
theory.  

The same approximations as above applied to the current-model leads to
the same reduced Maxwell-Bloch theory \eVIx, since $\d_t ( \psi 
e^{-i\omo t} ) \approx -i\omo \psi e^{-i\omo t} $.  

\bigskip
\noindent
{\it 6.2  Exact One-Polariton States}
\medskip

\def\st{\tilde{S}}

One can construct the one-polariton state exactly for the above
model.  Let us define momentum space S-operators as follows:
\eqn\eVIxii{\eqalign{
\st^\pm (k) &= \int \frac{dx}{\sqrt{2\pi}} ~ e^{\pm i k x} S^\pm (x, 0) \cr
\st_3  (k) &= \int \frac{dx}{\sqrt{2\pi}} ~ e^{- i k x} S_3  (x, 0) \cr
. }}
The hamiltonian now reads 
\eqn\eVIxiii{
H = \int dk_e ~ k_e ~ \ah^\dagger (k_e) \ah (k_e) 
-i \frac{\beta}{2} \sqrt{\frac{\omo}{2}} \int dk 
\( \ah (k) \st^+ (k) - \ah^\dagger (k) \st^- (k) \) . }
The state $\ovac$, defined in \eIIxxxvib, is now, in contrast to 
before, the exact ground state:
\eqn\eVIxiv{
H\ovac  = 0. }
By dropping $\hatom$ in \eVIxiii, we have merely shifted the 
ground state energy by $ N\omo/2 $. 

Below, we will need $\st_3 (k) \ovac$.  From \eVIxii\ and \eIIxxxviii,
one has 
\eqn\eVIxv{
\st_3 (k) \ovac = - \nl \sqrt{2\pi} \delta_L (k) \ovac, }
where $\delta_L$ is a delta function $\int dk \delta_L (k) = 1$,
and 
\eqn\eVIxvi{
\lim_{L\to \infty} \frac{\delta_L (0) }{L} = \inv{2\pi} .}
In this way, 
\eqn\eVIxvii{
\hatom \ovac = \sqrt{2\pi} \frac{\omo}2 \st_3 (0) \ovac 
= - \frac{N\omo}{2} \ovac . }

Understanding that the polariton quasiparticle is a combination
of photon and atomic degrees of freedom, let us take as an ansatz
for the one-polariton states:
\eqn\eVIxviii{
\ket{k_e} = \( \ah^\dagger (k_e) - i \lambda (k_e) \st^+ (k_e ) \) 
\ovac , }
where $\lambda$ is some function of $k_e$. 
One has 
\eqn\eVIxix{\eqalign{
\[H, \ah^\dagger (k_e) \] &= k_e ~ \ah^\dagger (k_e) - i \frac{\beta}2 
\sqrt{ \frac {\omo}2 } \st^+ (k_e )  \cr 
\[H, \st^+ (k_e )\] &= -i \frac{\beta}{2}  \sqrt{ \frac{\omo}{4\pi} }
\int dk' ~ \ah^\dagger (k') \st_3 (k' - k_e ) . \cr}}
This gives
\eqn\eVIxx{
H \ket{k_e} = 
\[ \( k_e + \frac{\beta}2 \sqrt{ \frac{\omo}2 } \nl \lambda (k_e) \) 
\ah^\dagger (k_e )  - i 
\frac{\beta}2 \sqrt{ \frac{\omo}2 } 
\st^+ (k_e ) 
\] \ovac .}
Thus, $\ket{k_e}$  is an exact eigenstate of $H$, 
\eqn\eVIxxi{
H \ket{k_e}  = \om_e (k_e) \ket{k_e} , }
with 
\eqn\eVIxxii{
\om_e = 
\frac{\beta}2 \sqrt{ \frac{\omo}2 } \inv{\lambda} , }
if $k_e$ satisfies
\eqn\eVIxxiii{
k_e = 
\frac{\beta}2 \sqrt{ \frac{\omo}2 } \( \inv{\lambda} - \nl \lambda \).}

One can thus view $\lambda$ as a spectral parameter, parameterizing the
envelope energy and momentum $\om_e, k_e$.  
Eliminating $\lambda$ one finds the dispersion relation
\eqn\eVIxxiv{
\om_e^2 - \om_e k_e =  m^2 , }
where $m^2$ is defined in \eIVxxiiib.  
The above dispersion relation is {\it exact}; thus one sees that
there are no one-loop corrections of the kind computed in section 4. 
This also means that the reduced Maxwell-Bloch theory does not incorporate
spontaneous emission effects. 

The result \eVIxxiv\ can now be compared with the result \eIVxxv\ 
obtained to lowest order in perturbation theory. 
Recalling that $\om_e$ and $k_e$ denote envelope quantities, we
let $\om = \om_0 + \om_e$, $ k = \omo + k_e$ and substitute in
\eIVxxv.   One obtains
\eqn\eVIxxv{
2 \omo \om_e + \om_e^2 - 2\omo k_e - k_e^2 
= 4 m^2 \frac{\omo^2}{ 2 \om_e \omo + \om_e^2 } . }
Using the slowly varying envelope inequalities, 
\eqn\slow{
\om_e^2 \ll \omo^2, ~~~~~ k_e^2 \ll \omo^2 ,}
one obtains precisely
\eVIxxiv. 

Remarkably, it is known that the model \eVIxi\ is integrable.
The Heisenberg operator equations of motion are 
\eqn\eVIxxvi{\eqalign{
(\d_t + \d_x ) \psi &= \frac{\beta}{4} S^- , ~~~~~~~
(\d_t + \d_x ) \psi^\dagger  = \frac{\beta}{4} S^+ \cr
\d_t S_3 &= - \beta \omo \( \psi^\dagger S^- + \psi S^+ \) \cr
\d_t S^+ &= \frac{\beta \omo}2 \psi^\dagger S_3 , ~~~~~~~
\d_t S^- = \frac{\beta \omo}2 \psi S_3 . \cr  }} 
These equations of motion have a zero-curvature representation 
\eqn\eVIxxvii{
[ \d_t + A_t , \d_x + A_x ] = 0, }
where $A_x, A_t$ are auxiliary $2\times 2$ matrices of
quantum operators:
\eqn\eVIxxviii{\eqalign{
A_x &= \mu \left(\matrix{ b_2^\dagger b_2 & S^+ \cr
S^- & b_1^\dagger b_1 \cr}\right) - \inv{\mu} \frac{\beta^2 \omo}{16} 
\left(\matrix{1&0\cr 0 &-1\cr } \right)  
+ \frac{\beta \omo}{2}  
\left(\matrix{0& - \psi^\dagger \cr \psi & 0 \cr}\right) 
\cr
A_t &=  \frac{\beta\omo}2 \left(\matrix{0& \psi^\dagger \cr - \psi & 0 \cr}
\right)  + \inv{\mu} \frac{\beta^2 \omo}{16} 
\left(\matrix{1&0\cr 0 &-1\cr } \right) 
. \cr}}
Above, $\mu$ is an arbitrary spectral parameter, and requiring \eVIxxvii\
to be valid for all $\mu$ is equivalent to \eVIxxvi. 

The zero-curvature representation allows the model to be solved
by the Quantum Inverse Scattering Method\ref\qism{V. E. Korepin,
N. M. Bogoliubov and A. G. Izergin, {\it Quantum Inverse Scattering
Method and Correlation Functions}, Cambridge University Press, 1993.}, 
as was carried out 
by Rupasov\ref\rrup{V. I. Rupasov, JETP Lett. 36 (1982) 142.}. 
The integrability 
leads to a Bethe-ansatz
construction of the multi-particle states that generalizes the above 
construction
 for the one-polariton states.  

\bigskip
\newsec{Semi-Classical Analysis}
\medskip

The reduced Maxwell-Bloch equations \eVIxxvi\ have been extensively
studied in a semi-classical approximation.  See e.g. \rhahn\eberly\
 and \ref\rlamb{G. L. Lamb, Rev. Mod. Phys. 43 (1971) 99.}. 
The nature of this semi-classical approximation is the following. 
Consider the expectation of the equations \eVIxxvi\ in the state
$\ovac$.  Let us assume that the atomic and photon correlators are
approximately decoupled:
\eqn\eVIxxix{
\vev{ \psi S^+ } \approx \vev{\psi} \vev{S^+} ,
}
where $\vev{O} = \lovac O \ovac$.  
In this semi-classical approximation $\vev{\psi}$ is now interpreted
as a classical electromagnetic field.   Define 
\eqn\eVIxxx{
\rho^\pm = S^+ \pm S^- .}
Imposing a reality condition $\vev{\psi} = \vev{\psi^\dagger}$, one obtains
the c-number equations
\eqn\eVIxxxi{\eqalign{
\d_t \vev{\rho^+}  &= \beta \omo~  
\vev{\psi} \vev{S_3}  \cr
\d_t \vev{S_3 }  &=  - \beta \omo ~ 
 \vev{\psi}  \vev{\rho^+}  \cr
\d_t \vev{\rho^-} &= 0. \cr}}
Since $\vev{\rho^- (t=0)} = 0$, the last equation above allows us to 
impose $\vev{\rho^-} = 0$ for all times.  

As a consequence of $\vec{S}^2$ being a Casimir for su(2), 
$\vev{\vec{S}} \cdot \vev{\vec{S}} $ is a constant of the motion. 
Having set $\vev{\rho^-}$ to zero, this implies
\eqn\eVIxxxii{
\vev{S_3}^2 + \vev{\rho^+}^2 = {\rm constant} = \( \nl \)^2 .  }
This constraint can be parameterized by introducing an angle function
$\Theta (x,t)$:
\eqn\eVIxxxiii{
\vev{S_3 (x,t)} = - \nl ~ \cos(\beta \Theta (x,t)), ~~~~~
\vev{\rho^+  (x,t)} = - \nl ~ \sin (\beta \Theta (x,t)) . }
The equations \eVIxxxi\ then imply
\eqn\eVIxxxiv{
\d_t \Theta = \omo  \vev{\psi}  . }
Inserting this into \eVIxxvi\ one gets the sine-Gordon like equation:
\eqn\eVIxxxv{
\( \d_t^2 + \d_t \d_x \) \Theta = - \frac{m^2 }{\beta } \nl 
\sin (\beta \Theta ) . }
The classical soliton solutions to this equation were observed
some time ago by McCall and Hahn\rhahn.

The sine-Gordon (SG) equation  is easily 
seen to be consistent with the one-polariton
dispersion relation obtained above in perturbation theory. 
Taking $\beta$ to be very small,
and expanding the $\sin (\beta \Theta)$ leads to the linear equation
\eqn\elinear{
\( \d_t^2 + \d_t \d_x \) \Theta = - m^2 ~ \Theta, }
with a dispersion relation that is precisely \eVIxxiv.  

The classical SG equation has a rich spectrum of solutions
consisting of solitons and breathers.  The lowest energy breather
solution can be identified with the polariton, as \elinear\ shows. 
The existence of these solutions suggests that the quantum
Maxwell-Bloch theory may have a rich spectrum of bound states
in addition to the polariton.  In the next section we attempt
to study this question by considering the quantum version of the
SG theory. 

\bigskip
\newsec{Quantum Sine-Gordon as an Effective Theory}

\medskip

In the last section we saw how the classical sine-Gordon equation
emerged from the reduced Maxwell-Bloch theory in a semi-classical
approximation wherein the electromagnetic field was treated
classically. 
Suppose one attempts to re-quantize the semi-classical treatment
by quantizing the sine-Gordon theory in the canonical manner.
What such a quantum theory has to do with the fully quantum
Maxwell-Bloch theory is a delicate question.  One can hope that
an intricate manifestation of the correspondence principle
in the end will save the day.  In this section we explore these
issues and conclude that the quantum sine-Gordon theory has
some validity.  

\def\xt{{\tilde{x}}}
\def\tt{{\tilde{t}}}

In order to make use of the standard quantization of sine-Gordon,
let us make a change of variables 
\eqn\eVIIIi{
\xt = 2x -t , ~~~~~\tt = t , }
such that 
\eqn\eVIIIii{
\d_t^2 + \d_t \d_x = \d_\tt^2 - \d_\xt^2 . }
The action which leads to the equations of motion \eVIxxxv\ is
\eqn\eVIIIiii{
S = \gamma \int d\tt d\xt ~ \[ \inv{2} 
\( \d_\tt \Theta \d_\tt \Theta - \d_\xt \Theta \d_\xt \Theta \) 
+ \frac{m^2}{ \beta^2} \cos (\beta \Theta ) \]  , }
where $\gamma$ is an arbitrary constant. 
In the classical theory the constant $\gamma$ is irrelevant, i.e.
the classical equations of motion are independent of 
$\gamma$.  In the quantum theory however, $\gamma$ determines the
fundamental commutation relations:
\eqn\eVIIIiv{
\[ \Theta (\xt , \tt ) , \d_\tt \Theta (\xt' , \tt ) \] = \frac{i}{\gamma}
\delta (\xt - \xt' ) , }
and is thus meaningful. 

To promote $\Theta$ to an operator and impose the commutation
relation \eVIIIiv\ is potentially perilous given the origin of
$\Theta$, i.e. as a way of solving the c-number constraint 
\eVIxxxii.  Let us try and interpret the quantization of $\Theta$ by
replacing \eVIxxxiii, \eVIxxxiv\ with {\rm operator} equations:
\eqn\eVIIIv{
S_3 = - \nl \cos (\beta \Theta), ~~~~~~~
\rho^+ = - \nl \sin (\beta \Theta)}
\eqn\eVIIIvi{
\d_t \Theta = \omo \> \psi . }
The S-commutation relations \eIIxxiv, after setting $\rho^- =0$, 
imply 
$[S_3 (x,t), \rho^+ (x',t')] = 0$.  This is consistent since
$[\Theta (x,t), \Theta (x', t')] = 0$.  

In \ref\lec{A. Leclair, Nucl. Phys. B 450 (1995) 753, hep-th/9505086.},
we argued that to connect with the quantum Maxwell-Bloch theory, one must
take $\gamma = 1$\foot{The discrepancy of $4\pi$ between equation
3.4 of \lec\ and equation \eIIxiv\ is because Heaviside-Lorentz
units were not used throughout in \lec; in particular the 
$4\pi$ in \eIIvii\ was omitted in \lec, which amounts to
a redefinition of $e$. All formulas in this paper are in the
esu system of units.}.  Let us repeat a version of this argument here. 
First, note that the operator equation \eVIIIvi\ combined with
the commutation relation \eVIix\ does not by itself give 
a commutation relation of the kind \eVIIIiv. 
In a sense, one must go to next to leading order in the slowly varying
envelope approximation in order to obtain \eVIIIiv, as follows. 
Recalling the reality condition imposed in the previous section
$\vev{\psi} = \vev{\psi^\dagger}$, let us impose this classically.
Then $\phi = 2 \psi \cos (\omo (t-x))$. Since $\d_t \phi 
\approx - 2 \omo \psi \sin (\omo (t-x))$, we have
$\d_t \phi \approx -2 \d_t \Theta \sin (\omo (t-x))$.  The term
in the action \eIIx\ for $\phi$ that determines the commutation
relations is 
\eqn\eVIIIvii{
\int dx dt ~ \inv{2} \d_t \phi \d_t \phi \approx
4 \int dx dt ~ \d_t \Theta \d_t \Theta \sin^2 ( \omo(t-x)) .}
Averaging over the rapid oscillations let us replace 
$\sin^2(\omo(t-x))$ by $1/2$.  Then the commutation relations
which follow from \eVIIIvii\ are 
\eqn\eVIIIviii{
\[ \Theta (x,t), \d_t \Theta (x', t)\] = \frac{i}{2} 
\delta(x-x') . }
Since $\delta(x-x') = 2 \delta (\xt - \xt')$ at equal time,
\eVIIIviii\ is precisely \eVIIIiv\ with $\gamma=1$. 

Further evidence for the relevance of the quantum SG theory defined
by \eVIIIiii\ with $\gamma = 1$ comes from the perturbative computations
in section 6, in particular the beta-function \eViv.  
From the 2-point function 
\eqn\eVIIIix{
\lvac \Theta (\xt, \tt) \Theta (0) \rvac_{m = 0}
= - \inv{4\pi} \log ( \tt^2 - \xt^2 ), } 
one has 
\eqn\eVIIIx{
\lvac e^{i\beta \Theta (\xt , \tt)} e^{-i\beta \Theta (0)} \rvac_{m = 0} 
= \inv{ (\tt^2 - \xt^2 )^{\beta^2 /4\pi }  }. }
This implies that the $\cos (\beta \Theta )$ operator in \eVIIIiii\
has anomalous mass dimension $\beta^2 / 4 \pi $.  Since the action 
$S$ is dimensionless, $m^2$ has a dimension of $2 - \beta^2 /4\pi$.  Since
$m^2 \propto \omo$, then this precisely corresponds to the beta-function
\eViv.  

\def\omt{\tilde{\om}}
\def\kt{\tilde{k}}
\def\mut{\tilde{\mu}}

One does not expect of course that the quantum SG theory precisely
reproduces the quantum corrections computed in section 4 for the
fully quantum Maxwell-Bloch theory.  One can study this explicitly
by computing one loop corrections in the SG model. Expanding out
the $\cos(\beta \Theta)$, the lagrangian is 
\eqn\eVIIIxi{
\CL = \inv{2} \tilde{\d}_\mu \Theta \tilde{\d}^\mu \Theta 
-  \frac{m^2}{2} \Theta^2 + \frac{m^2 \beta^2}{24} ~ \Theta^4 + \ldots }
The one loop contribution to the photon self-energy can be computed
using standard perturbation theory.  In the conventional coordinates
$\xt, \tt$, the frequency and wave-vector $\tilde{\om} , \tilde{k}$
are related to the envelope quantities by $\omt = \om_e - k_e /2$,
$\kt = k/2$, so that $\omt^2 - \kt^2 = \om_e^2 - \om_e k_e$. Introducing
a u.v. cuttoff $\tilde{\mu}$ as in section 4, one finds the 
dispersion relation:
\eqn\eVIIIxii{
\om_e^2 - \om_e k_e - m^2 
\[ 1 - \frac{\beta^2}{8\pi} \( \log \( \frac{\mut}{m} 
+ \sqrt{ \frac{\mut^2}{m^2} - 1 } \)^2  - i\pi \) \] =0. }
This should be compared to the dispersion relation 
$\om^2 - k^2 - \Sigma = 0$ for the current-model near resonance,
i.e. with $\om = \omo + \om_e$, $k = \omo + \om_e$. 
Using the slowly varying inequalities \slow, and the expression 
\eIVxxviii, one obtains:
\eqn\eVIIIxiii{
\om_e^2 - \om_e k_e - m^2 
\[ 1 + \frac{\beta^2}{4 \pi} - \frac{\beta^2}{8\pi} 
\frac{\omo}{\om_e} \( \log \( \frac{\mu^2}{\omo^2} -1 \)  
 - i\pi \) \] =0. }
The expressions \eVIIIxiii\
and \eVIIIxii, including the imaginary parts, 
agree when both cutoffs are large  
and when $\om_e \approx \omo$; however this  
contradicts the slowly varying inequalities.  

The conclusion of the above analysis is that the quantum SG theory
captures some aspects of the fully quantum Maxwell-Bloch theory,
in particular the current-model defined in section 2, and does
incorporate spontaneous emission,  but is 
not equivalent to it even in the slowly varying envelope approximation. 

As a step toward understanding the spectrum of the quantum
Maxwell-Bloch theory, one can assume the approximate validity
of the quantum SG description. 
 The quantum
SG spectrum is known to consist of breathers, the lowest mass
breather being the particle associated with the SG field $\Theta$
itself, and a pair of solitons.  See e.g. \ref\rzz{A. B. Zamolodchikov
and Al. B. Zamolodchikov, Ann. Phys. 120, (1979) 253.}. 
The mass of the n-th breather is given by 
\eqn\esga{
m_n = 2 m_s \sin \( \frac{n \xi }{16} \), ~~~~~~n=1,2,...< 
\frac{8\pi}{\xi }, ~~~~\xi = \frac{\beta^2}{1-\beta^2 /8 \pi} .}
where $m_s$ is the mass of the soliton. 
As $\beta \to 0$, the mass of the lowest breather $m_1$ approaches $m$
thus the polariton is identified as the lowest breather.  The higher
breathers are polariton bound states.  
As $\beta \to 0$, $m_s \approx 8 m/\beta^2$, thus for very small
$\beta$, the mass of the soliton can be very large compared with the 
polaritons.  In the quantum theory, the polariton can actually be
viewed as a bound state of two solitons.  

It would be very interesting
to understand whether aspects of this spectrum can be seen in the models
defined in section 2.

\bigskip
\newsec{Conclusions} 

We have defined some models which describe quantized radiation in
interaction with a medium of two-level atoms arranged in a fiber
geometry.  Our main computational results are the photon self-energy
\eIVxxiii\ and \eIVxxviii, which determine the first quantum corrections
to the polariton dispersion relation.   We also compared our results
with known semi-classical results in the slowly-varying envelope
and rotating-wave approximations, and argued for the approximate
validity of
the quantum sine-Gordon theory. 
We found that the model which follows from an interaction hamiltonian
$-\vec{d} \cdot \vec{E}$ (current-model) 
is better behaved than the model which
follows from the minimal coupling $\vec{p} \to \vec{p} - e \vec{A} $,
the latter suffering from infrared divergences.  
In the current model we derived a renormalization group equation
for the energy splitting of the two-level atoms which follows from
the beta function \eViv, and shows that the splitting is screened
at higher energies.  

Though the quantum corrections are generically small, we hope that
the trend toward fabricating smaller optical devices will eventually
lead to the observation of these quantum effects. 

The models defined in section 2 deserve further theoretical study,
in particular it would be interesting to determine whether they
have a bound state spectrum that resembles the spectrum of the
quantum sine-Gordon
theory.

\bigskip
\bigskip

\centerline{\bf Acknowlegements} 

I would like to thank A. Gaeta, B. Gerganov, R. Konik, S. Herrandies, 
P. Lepage, 
S. Lukyanov, and H. Saleur for discussions.  
This work is supported by the National Science Foundation, in part through
the National Young Investigator program. 

\listrefs
\end